\documentclass[preprint,12pt,3p]{elsarticle}

\usepackage{float}

\usepackage{anyfontsize}

\usepackage{amssymb}
\usepackage{amsthm}
\usepackage{amsmath}

\usepackage[figuresright]{rotating}

\usepackage{subcaption}

\usepackage{graphicx}
\usepackage{dcolumn}
\usepackage{bm}
\usepackage{epstopdf}
\usepackage{epsfig}
\usepackage[colorlinks = true,
linkcolor = blue,
urlcolor  = blue,
citecolor = blue,
anchorcolor = blue]{hyperref}
\usepackage{url}

\newtheorem{theorem}{Theorem}

\theoremstyle{remark}
\newtheorem{remark}[theorem]{Remark}

\begin{document}
	
	\begin{frontmatter}
		
		\title{Spontaneous symmetry-breaking in the nonlinear Schr\"odinger equation on star graphs with inhomogeneities}
		
		
        \author[au1]{Rahmi Rusin}
		\author[au2]{Hadi Susanto}		

        \address[au1]{Department of Mathematics, Faculty of Mathematics and Natural Sciences, Universitas Indonesia,\\ Gedung D Lt.\ 2 FMIPA Kampus UI Depok, 16424, Indonesia}
		\address[au2]{Department of Mathematics, Khalifa University, PO Box 127788, Abu Dhabi, United Arab Emirates}


		\begin{abstract}
We investigate the nonlinear Schr\"odinger equation on a three-edge star graph, where each edge contains a linear localized inhomogeneity in the form of a Dirac delta linear potential. Such systems are of significant interest in studying wave propagation in networked structures, with applications in, e.g., Josephson junctions. By reducing the system to a set of finite-dimensional coupled ordinary differential equations, we derive explicit conditions for the occurrence of a symmetry-breaking bifurcation in a symmetric family of solutions. This bifurcation is shown to be of the transcritical type, and we provide a precise estimate of the bifurcation point as the propagation constant, which is directly related to the solution norm, is varied. In addition to the symmetric states, we explore non-positive definite states that bifurcate from the linear solutions of the system. These states exhibit distinct characteristics and are crucial in understanding solutions of the nonlinear system. Furthermore, we analyze the typical dynamics of unstable solutions, showing their behavior and evolution over time. Our results contribute to a deeper understanding of symmetry-breaking phenomena in nonlinear systems on metric graphs and provide insights into the stability and dynamics of such solutions.
		\end{abstract}

	\end{frontmatter}
	
	

\section{Introduction}

The ground state of physical systems typically inherits the symmetry of the external potential acting on the physical field or wave function. However, this rule may only hold in the weakly nonlinear regime in the presence of nonlinearity. As the strength of the nonlinearity increases, spontaneous symmetry breaking can occur, leading to a scenario where symmetric wave functions no longer represent the ground state. In such cases, the symmetric solutions become unstable against non-symmetric perturbations, often through a pitchfork bifurcation.

The concept of spontaneous symmetry breaking in the context of the nonlinear Schr\"odinger equation was likely first introduced by Davies \cite{davies1979symmetry} in a model describing the interactions of quantum particles through a three-dimensional isotropic potential. In this work, the symmetry breaking was characterized as a bifurcation involving the loss of rotational symmetry in the ground state. A simpler model, formulated as a system of coupled ordinary differential equations and exhibiting symmetry-breaking bifurcation, was later presented in \cite{eilbeck1985discrete}. Since then, the phenomenon of symmetry breaking has been extensively studied in a wide range of contexts, including Bose-Einstein condensates \cite{albiez2005direct,zibold2010classical}, metamaterials \cite{liu2014spontaneous}, spatiotemporal complexity in lasers \cite{green1990spontaneous}, photorefractive media \cite{kevrekidis2005spontaneous}, biological slime molds \cite{sawai2000spontaneous}, coupled semiconductor lasers \cite{heil2001chaos}, and nanolasers \cite{hamel2015spontaneous}. In these systems, the breaking of inversion symmetry in a double-well potential manifests as a transition to two states localized in one of the potential wells, which are mirror images of each other.

Theoretical studies on spontaneous symmetry breaking bifurcations have also explored systems of linearly coupled nonlinear Schr\"odinger equations, which admit asymmetric two-component soliton modes \cite{wright1989solitary,pare1990approximate,mauimistov1991propagation,akhmediev1993novel,malomed1996symmetric}. Other works have investigated unstable linearly coupled dark solitons leading to bosonic Josephson vortices \cite{qadir2012fluxon,su2013kibble,qadir2014multiple} and symmetry breaking of linearly coupled vortices \cite{salasnich2011spontaneous,chen2017spontaneous}. For a comprehensive overview of the subject, the reader is referred to the book \cite{malomed2013spontaneous}.

In this paper, we consider a novel system by considering the nonlinear Schr\"odinger equation on a three-edge star graph, where each edge contains a localized inhomogeneity in the form of a Dirac delta potential. We study the symmetry-breaking bifurcation in this system, representing a metric graph—a network-shaped structure consisting of vertices connected by edges. The Schr\"odinger equation is defined on the edges with appropriate boundary conditions at the vertices, making it a suitable model for wave propagation in systems analogous to a thin neighborhood of a graph. This framework has gained attention recently due to its potential as a paradigm model for exploring topological effects in nonlinear wave propagation, as reviewed in \cite{noja2014nonlinear}. Our study is particularly relevant in the context of multi-edge Josephson junctions \cite{susanto2019soliton}. A Josephson junction is a quantum structure composed of two superconducting electrodes separated by a thin insulating barrier. When three semi-infinite Josephson junctions are arranged such that their ends converge at a single common point, they form a structure known as a tricrystal junction. Such junctions have been experimentally fabricated and utilized as a tool to investigate the symmetry properties of the order parameter in high-temperature superconductors, as documented in \cite{tsuei1994pairing,miller1995use,tsuei2000phase,tsuei2000pairing}. Beyond tricrystal configurations, researchers have also explored tetracrystal junctions, which consist of star-shaped graphs with four arms. These structures have been extensively studied in experimental settings in \cite{tsuei2000pairing,tomaschko2012phase}. Point-like inhomogeneities along a Josephson junction have also been created experimentally; see \cite{golubov1988dynamics, vystavkin1988first, cirillo1985inductively}. Our present work introduces a model of multi-arm systems containing point (Dirac delta) inhomogeneities.

While the real line can be viewed as a two-edge star graph with two-fold symmetry, three-edge star graphs exhibit rotational symmetry of order three. A striking difference between the nonlinear Schr\"odinger equation on the real line and on three-edge star graphs is that the latter, under Kirchhoff conditions at the vertex, does not admit a unique "trapped soliton" state as the ground state \cite{adami2012structure}. In this paper, we introduce another notable distinction by incorporating a linear Dirac delta potential on each arm of the star graph. We report that the symmetry-breaking bifurcation in this system occurs through a transcritical bifurcation, which stands in contrast to the standard symmetry-breaking driven by a pitchfork bifurcation \cite{snyder1991physics,yang2013stability}. While transcritical bifurcations have been observed in double-well potentials, they have only been reported in asymmetric cases \cite{yang2012classification,yang2013stabilityswitch}. The stability switching between the involved branches in our system is novel and does not align with any of the previously reported cases in \cite{yang2013stabilityswitch}. Additionally, we identify a critical distance of the external potential minima from the vertex below, where no symmetry breaking occurs.

Although our system can be interpreted as a Schr\"odinger equation with a triple-well potential, it fundamentally differs from the case on the real line \cite{kapitula2006three,goodman2011hamiltonian,goodman2017bifurcations}, where the presence of a third well results in all bifurcations being of the saddle-node type. This highlights the unique behavior of the system on a three-edge star graph.

The paper is structured as follows. Section \ref{model} introduces the mathematical model and its key components. We then discuss the underlying linear states of the system in Section \ref{linear}. We derive a transcendental equation that determines the bifurcation points of eigenstates from the zero solution. Using the coupled mode reduction method, we reduce the problem to a finite-dimensional dynamical system and analyze the existence and stability of standing localized solutions of the nonlinear equation. In Section \ref{stability}, we investigate the stability of the states. We demonstrate the existence of a threshold point at which symmetric states become unstable and discuss the critical distance of the external potential minima from the vertex below which symmetric states remain stable. In Section \ref{num_ori}, we discuss our methods for solving the original differential equations. In Section \ref{num}, we present our findings from solving the coupled-mode and differential equations. We illustrate the typical dynamics of the standing waves when unstable using numerical simulations. Finally, in Section \ref{concl}, we summarize our findings and discuss their implications.

\section{Mathematical model}
\label{model}

Our domain is a graph \( G \) consisting of three semi-infinite lines connected at a common vertex. The Schr\"odinger equation is formulated on the Hilbert space \( L^2(G) = \bigoplus_{k=1}^3 L^2(\mathbb{R}^+) \). The wave function along each semi-infinite line is described by the following nonlinear Schr\"odinger equation
\begin{equation}
iu^{(k)}_t = -u^{(k)}_{xx} - |u^{(k)}|^2 u^{(k)} + \omega u^{(k)} - \delta(x - a) u^{(k)},
\label{nls}
\end{equation}
where the superscripts \( k = 1, 2, 3 \) label the different branches of the system, and the subscripts denote derivatives with respect to the variables. The Dirac delta potential $\delta$ represents a localized inhomogeneity located at $x=a$. At the vertex \( x = 0 \), we impose the free Kirchhoff boundary conditions
\begin{equation}
\label{bc}
\begin{aligned}
\sum_{k=1}^3 u^{(k)}_x(0) = 0, \quad u^{(1)}(0) = u^{(2)}(0) = u^{(3)}(0).
\end{aligned}
\end{equation}
The wave function \( u(x, t) = \bigoplus_{k=1}^3 u^{(k)}(x, t) \), where \( x, t \in \mathbb{R}^+ \), resides in the Sobolev space \( H^1(G) = \bigoplus_{k=1}^3 H^1(\mathbb{R}^+) \).

The system \eqref{nls} with boundary conditions \eqref{bc} conserves the squared \( L^2 \) norm \( N = \|u\|^2 = \langle u, u \rangle \), where the inner product is defined as
\begin{equation}
\label{innerprod}
\langle u, v \rangle = \sum_{k=1}^3 \int_{0}^{\infty} u^{(k)} \bar{v}^{(k)} \, dx.
\end{equation}
The quantity \( N \) is known as the optical power in the context of nonlinear optics, or the number of particles in the context of Bose-Einstein condensates.

The bound states of \eqref{nls} satisfy the equation
\begin{equation}
u^{(k)}_{xx} - \omega u^{(k)} + {u^{(k)}}^3 + \delta(x - a) u^{(k)} = 0.
\label{eqbs}
\end{equation}
Our objective is to study solutions of \eqref{eqbs}, which are equilibria of \eqref{nls}, and to determine their stability. To achieve this, we first employ a coupled mode reduction approach to \eqref{nls} by leveraging the eigenstates of the linear part of the system. For the nonlinear Schr\"odinger equation on the line with a double-well potential, it has been established that, on large but finite time scales, the dynamics are governed by a finite-dimensional dynamical system \cite{marzuola2010long,goodman2015self}. In this work, we assume that the results of \cite{marzuola2010long,goodman2015self} can be extended to our case, with a formal proof deferred to future research. We then complement our analysis with numerical solutions of the original nonlinear system.

\section{Coupled mode approximations}
\label{linear}

In this section, we derive a coupled mode approximation of the governing equation \eqref{nls}. We begin by determining the linear eigenstates of the system and then explain how to find solutions of \eqref{nls} as continuations of these eigenstates.

\subsection{Linear States}
In the limit $u \to 0$, Equation \eqref{eqbs} reduces to the linear system
\begin{equation} 
u^{(k)}_{xx} - \omega u^{(k)} + \delta(x - a) u^{(k)} = 0.
\label{eqbsl}
\end{equation}
This is equivalent to the linear system $u^{(k)}_{xx} - \omega u^{(k)} = 0$ for $x \neq a$, with the matching conditions
\begin{eqnarray}
u^{(k)}(a^+) = u^{(k)}(a^-), \quad u^{(k)}_x(a^+) - u^{(k)}_x(a^-) = -u^{(k)}(a).\label{match}
\end{eqnarray}
Additionally, at $x = 0$, the boundary conditions \eqref{bc} still hold.

The general solution of \eqref{eqbsl} is given by
\begin{equation}
u^{(k)} =
\left\{
\begin{array}{ll}
A^{(k)} e^{-\sqrt{\omega} (x - a)}, & x > a, \\
B^{(k)} e^{-\sqrt{\omega} (x - a)} + C^{(k)} e^{\sqrt{\omega} (x - a)}, & x < a.
\end{array}
\right.
\label{tamb}
\end{equation}
By applying the matching and boundary conditions, the function \eqref{tamb} is a solution of the linear system only if $\omega$ satisfies the transcendental relation
\begin{equation} 
\left(1 - \left(2 \sqrt{\omega} - 1\right) e^{2 a \sqrt{\omega}}\right) \left(1 + \left(2 \sqrt{\omega} - 1\right) e^{2 a \sqrt{\omega}}\right)^2 = 0.
\label{omega}
\end{equation}
This equation determines the bifurcation points of the linear states.

\begin{figure}[tbhp!]
	\centering
	\includegraphics[scale=0.55]{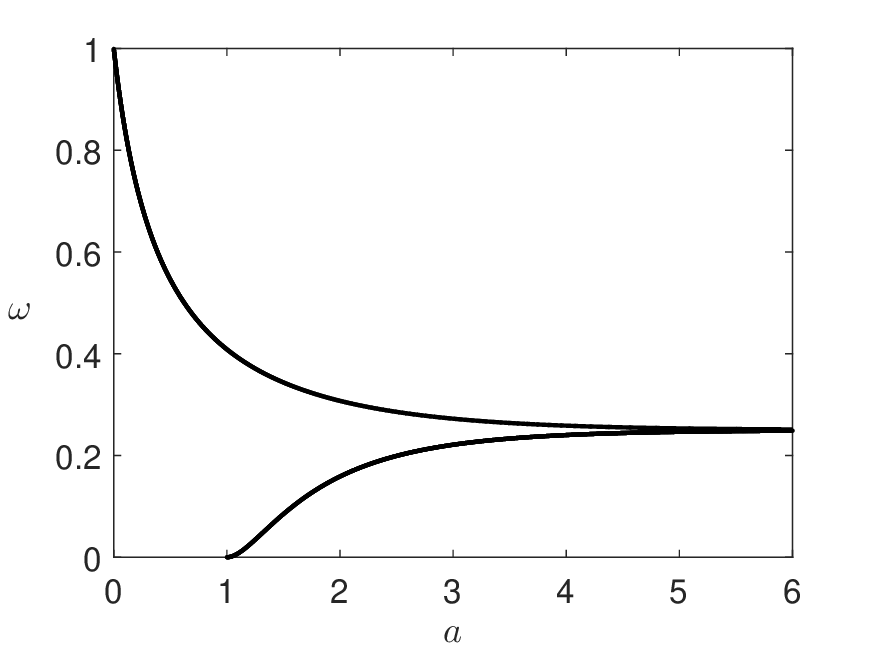}
	\caption{Eigenvalues as a function of $a$. The upper curve corresponds to $\omega_0$.} 
	\label{fig:a_vs_omega}
\end{figure}

Equation \eqref{omega} yields two eigenvalues, $\omega_0$ and $\omega_1$, where $\omega_1$ has multiplicity two. Specifically
\begin{equation}
a = -\frac{\ln \left(2 \sqrt{\omega_0} - 1\right)}{2 \sqrt{\omega_0}} \approx -\frac{1}{2}\left(\omega_0 - 1\right) + \frac{5}{8}\left(\omega_0 - 1\right)^2 - \frac{35}{48}\left(\omega_0 - 1\right)^3 + \cdots,
\label{o0}
\end{equation}
and
\begin{equation}
a = -\frac{\ln \left(1 - 2 \sqrt{\omega_1}\right)}{2 \sqrt{\omega_1}} \approx 1 + \sqrt{\omega_1} + \frac{4\omega_1}{3} + 2 \omega_1^{3/2} + \cdots.
\label{o1}
\end{equation}
We observe that the eigenfunction corresponding to $\omega_0$ exists for all values of $a$, while the eigenfunctions corresponding to $\omega_1$ exist only for $a \geq 1$. In the limit $a \to \infty$, both $\omega_0$ and $\omega_1$ approach $1/4$. The behavior of $\omega_0$ and $\omega_1$ as functions of $a$ is illustrated in Fig.~\ref{fig:a_vs_omega}.

\begin{figure}[tbhp!]
	\centering
	\includegraphics[scale=0.55]{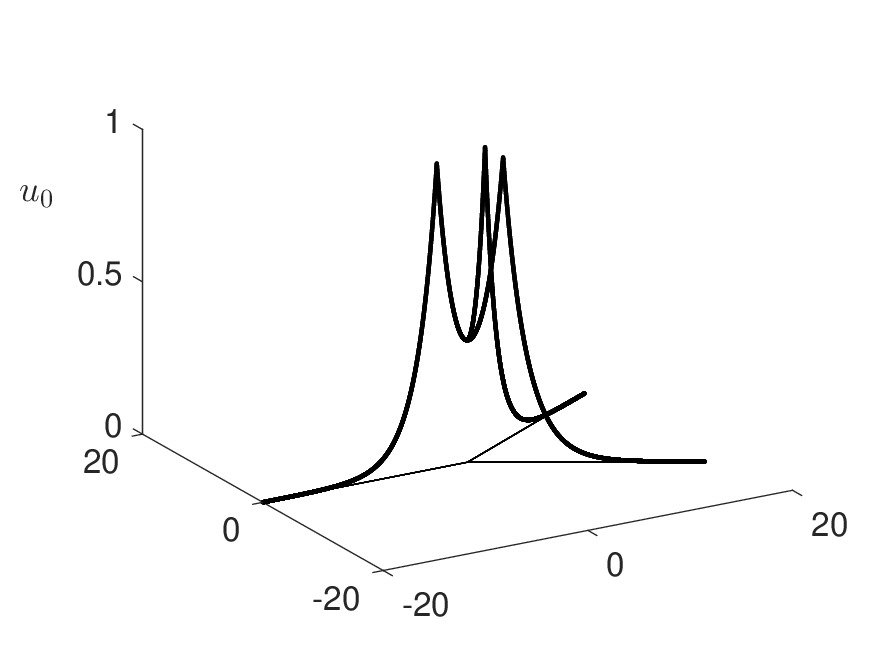}
	\includegraphics[scale=0.55]{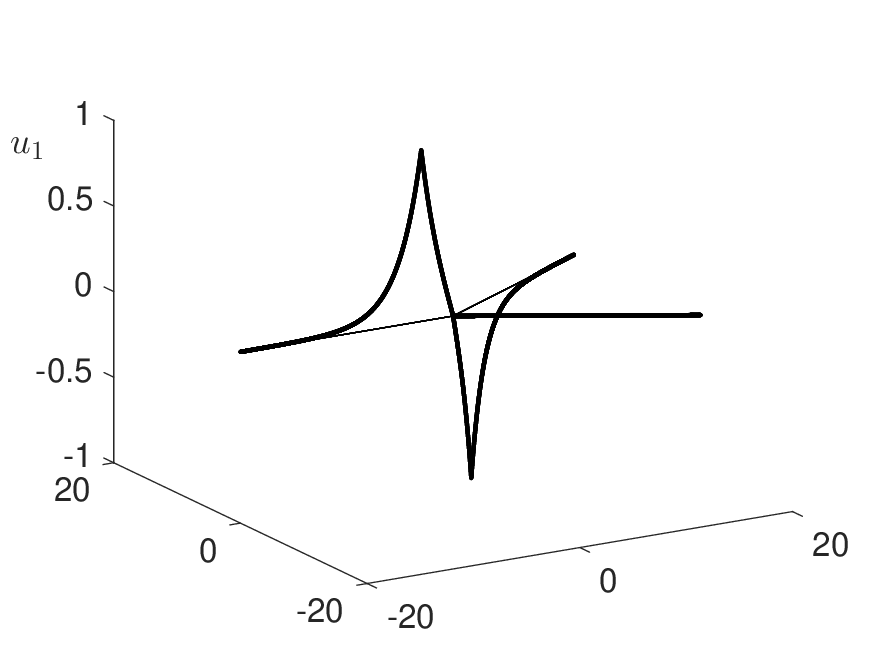}
	\includegraphics[scale=0.55]{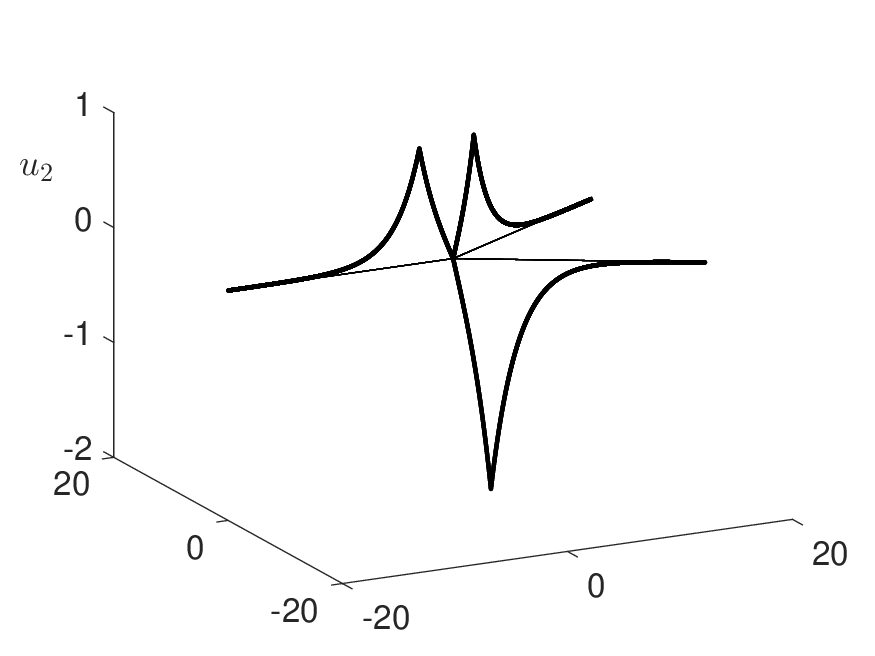}
	\caption{Plots of the eigenfunctions \eqref{efs} for $a = 3$.} 
	\label{fig:soln}
\end{figure}

Let $u_0(x)$ denote the eigenfunction corresponding to the eigenvalue $\omega_0$, and let the eigenspace corresponding to $\omega_1$ be spanned by the eigenfunctions $u_1(x)$ and $u_2(x)$. These eigenfunctions are given by
\begin{equation}
\begin{aligned}
u_0^{(1)} &= u_0^{(2)} = u_0^{(3)} = f(x; \omega_0) := 
\begin{cases}
e^{-\sqrt{\omega_0}(x - a)}, & x > a, \\
-\frac{1 - 2 \sqrt{\omega_0}}{2 \sqrt{\omega_0}} e^{-\sqrt{\omega_0}(x - a)} + \frac{1}{2 \sqrt{\omega_0}} e^{\sqrt{\omega_0}(x - a)}, & x < a,
\end{cases} \\
u_1^{(1)} &= -u_1^{(2)} = f(x; \omega_1), \quad u_1^{(3)} = 0, \\
u_2^{(1)} &= u_2^{(2)} = -\frac{u_2^{(3)}}{2} = f(x; \omega_1).
\end{aligned}
\label{efs}
\end{equation}
The eigenfunctions are plotted in Fig.~\ref{fig:soln}. Using the inner product defined in \eqref{innerprod}, it can be verified that $u_0$, $u_1$, and $u_2$ are mutually orthogonal.

\begin{remark}
\label{rem1}
The governing equation \eqref{eqbsl} is invariant under rotation (i.e., cyclic permutation of the arm indices). Due to this invariance, the following identities hold
\begin{eqnarray}
- u_1 \pm u_2 \, \textbf{`='} \, 2 u_1, \quad \pm 3 u_1 - u_2 \, \textbf{`='} \, 2 u_2,
\end{eqnarray}
where $\textbf{`='}$ denotes equality under cyclic permutation. This observation is crucial for the subsection below, where different sets of coefficients in the linear combination of the eigenfunctions may represent the same solution.
\end{remark}

\subsection{Formulation of the Finite Dimensional System}
\label{sec2}

In this subsection, we derive a finite-dimensional system from the governing equation \eqref{nls} using a coupled mode reduction method. This approach restricts the system to the bound state manifold, simplifying the analysis while retaining the essential dynamics.

\subsubsection{Ansatz and Projection}
Let the ansatz for solutions of the nonlinear equation \eqref{nls} be
\begin{equation} 
u(x) = c_0(t) \widehat{u}_0(x) + c_1(t) \widehat{u}_1(x) + c_2(t) \widehat{u}_2(x),
\label{ansatz}
\end{equation}
where $\widehat{u}_j$, $j = 0, 1, 2$, are the normalized eigenfunctions from \eqref{efs}. Substituting the ansatz into \eqref{nls} and noting that $\widehat{u}_j$ satisfies the linear equation \eqref{eqbsl} with the corresponding eigenvalue $\omega_j$ ($\omega_2 = \omega_1$), we obtain
\[
\begin{aligned}
i\left(\dot{c}_0 \widehat{u}_0 + \dot{c}_1 \widehat{u}_1 + \dot{c}_2 \widehat{u}_2\right) &= \left(\omega - \omega_0\right) c_0 \widehat{u}_0 + \left(\omega - \omega_1\right) c_1 \widehat{u}_1 + \left(\omega - \omega_1\right) c_2 \widehat{u}_2 \\
&\quad - \left(c_0 \widehat{u}_0 + c_1 \widehat{u}_1 + c_2 \widehat{u}_2\right)^2 \left(\bar{c}_0 \widehat{u}_0 + \bar{c}_1 \widehat{u}_1 + \bar{c}_2 \widehat{u}_2\right),
\end{aligned}
\]
where the overdot denotes the time derivative. 

Projecting this equation onto the eigenstates $\widehat{u}_j$ and defining the coefficients $g_{ijkl} = \langle \widehat{u}_i \widehat{u}_j \widehat{u}_k, \widehat{u}_l \rangle$, we derive the finite-dimensional dynamical system
\begin{equation} 
\begin{aligned}
i \dot{c}_0 &= \left(\omega - \omega_0\right) c_0 - g_{0000} |c_0|^2 c_0 - g_{0110} \left(\bar{c}_0 c_1^2 + 2 c_0 |c_1|^2\right) - g_{1120} \left(2 |c_1|^2 c_2 + c_1^2 \bar{c}_2\right) \\
&\quad - g_{0110} \left(\bar{c}_0 c_2^2 + 2 c_0 |c_2|^2\right) + g_{1120} |c_2|^2 c_2, \\
i \dot{c}_1 &= \left(\omega - \omega_1\right) c_1 - g_{0110} \left(2 |c_0|^2 c_1 + c_0^2 \bar{c}_1\right) - g_{1111} |c_1|^2 c_1 \\
&\quad - g_{1120} \left(2 \bar{c}_0 c_1 c_2 + 2 c_0 \bar{c}_1 c_2 + 2 c_0 c_1 \bar{c}_2\right) - g_{1221} \left(\bar{c}_1 c_2^2 + 2 c_1 |c_2|^2\right), \\
i \dot{c}_2 &= \left(\omega - \omega_1\right) c_2 - g_{1120} \left(\bar{c}_0 c_1^2 + 2 c_0 |c_1|^2\right) - g_{0110} \left(2 |c_0|^2 c_2 + c_0^2 \bar{c}_2\right) \\
&\quad - g_{1221} \left(2 |c_1|^2 c_2 + c_1^2 \bar{c}_2\right) + g_{1120} \left(\bar{c}_0 c_2^2 + 2 c_0 |c_2|^2\right) - g_{1111} |c_2|^2 c_2.
\end{aligned}
\label{sys1}
\end{equation}

\subsubsection{Approximation of Coefficients}
Before analyzing the equilibrium solutions of \eqref{sys1}, we approximate the values of the coefficients $g_{ijkl}$. For $a \gg 1$, the coefficients are approximately related by
\begin{equation} 
g_{0000} = g_{0110} = \sqrt{2} g_{1120} = \frac{2}{3} g_{1111} = 2 g_{1221}.
\label{appr}
\end{equation}
This approximation introduces an exponentially small error for large $a$ but significantly simplifies the analysis. 

Scaling the time as $t \rightarrow \Gamma t$, where $\Gamma = 1 / g_{0000}$, the system \eqref{sys1} becomes
\begin{equation} 
\begin{aligned}
i \dot{c}_0 &= \Gamma \left(\omega - \omega_0\right) c_0 - |c_0|^2 c_0 - \left(\bar{c}_0 c_1^2 + 2 c_0 |c_1|^2\right) - \frac{1}{\sqrt{2}} \left(\bar{c}_2c_1^2 +2 c_2|c_1|^2 \right) \\
&\quad - \left(\bar{c}_0 c_2^2 + 2 c_0 |c_2|^2\right) + \frac{1}{\sqrt{2}} |c_2|^2 c_2, \\
i \dot{c}_1 &= \Gamma \left(\omega - \omega_1\right) c_1 - \left(\bar{c}_1c_0^2+2c_1 |c_0|^2 \right) - \frac{1}{2} \left(\bar{c}_1 c_2^2 + 2 c_1 |c_2|^2\right)  \\
&\quad - \sqrt{2} \left(\bar{c}_0 c_1 c_2 + c_0 \bar{c}_1 c_2 + c_0 c_1 \bar{c}_2\right) - \frac{3}{2} |c_1|^2 c_1, \\
i \dot{c}_2 &= \Gamma \left(\omega - \omega_1\right) c_2 - \frac{1}{\sqrt{2}} \left(\bar{c}_0 c_1^2 + 2 c_0 |c_1|^2\right) - \left(\bar{c}_2c_0^2+2c_2 |c_0|^2\right) \\
&\quad - \frac{1}{2} \left(\bar{c}_2c_1^2 +2c_2 |c_1|^2\right) + \frac{1}{\sqrt{2}} \left(\bar{c}_0 c_2^2 + 2 c_0 |c_2|^2\right) - \frac{3}{2} |c_2|^2 c_2.
\end{aligned}
\label{sys2}
\end{equation}
In the limit $a \rightarrow \infty$, $\Gamma = 12$.

\begin{remark}\label{rem2}
The systems \eqref{sys1} and \eqref{sys2} are invariant under the transformation $c_1 \to -c_1$. This implies that if $(c_0, c_1, c_2)$ is a solution of the systems, then $(c_0, -c_1, c_2)$ is also a solution.
\end{remark}

\subsection{Equilibrium Solutions}
\label{eqm}

\begin{figure}[htbp!]
	\centering
   \includegraphics[clip=,scale=0.2]{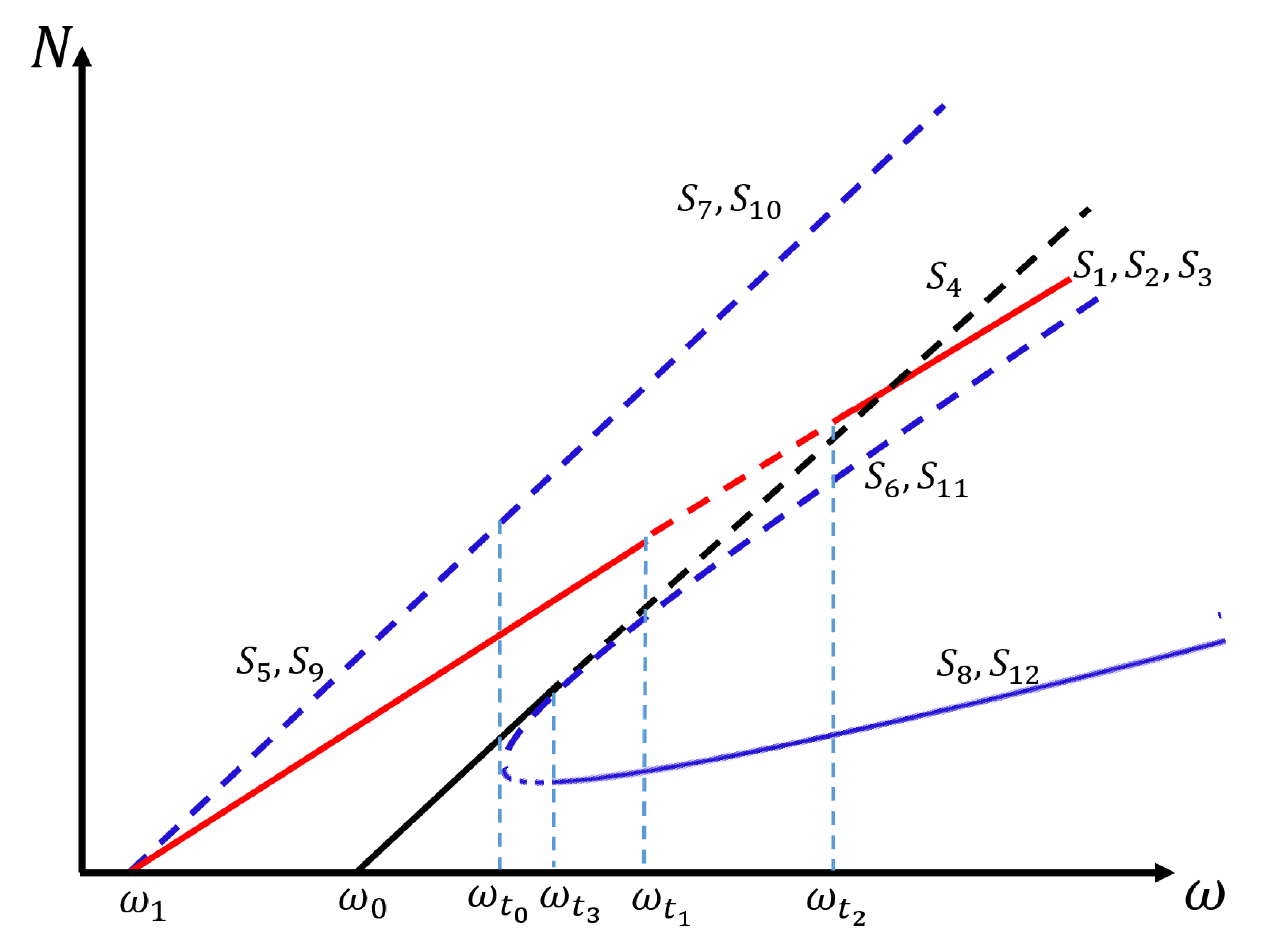}
	\caption{Sketch of the bifurcation diagram in the $(\omega, N)$-plane for the various equilibrium solutions discussed in Section \ref{eqm} for $a \gg 1$. This is an unscaled plot for clarity. See the text for the meaning of the labels.}
	\label{fig:sketch}
\end{figure}

\begin{figure}[htbp!]
	\centering
	\subcaptionbox{\label{subfig:solABC_u1u2}}{\includegraphics[scale=0.55]{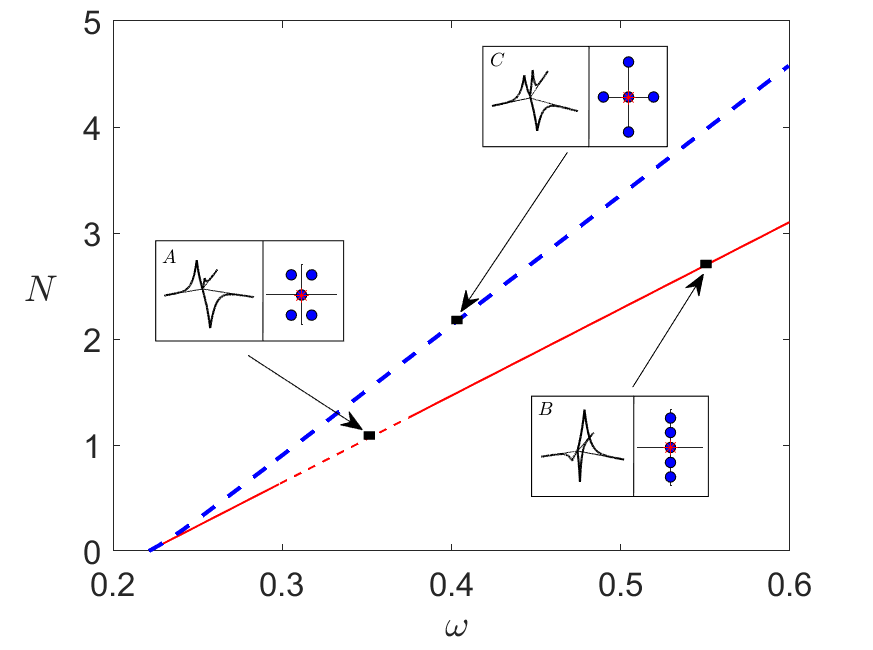}}
	\subcaptionbox{\label{subfig:solABC_u0}}{\includegraphics[scale=0.55]{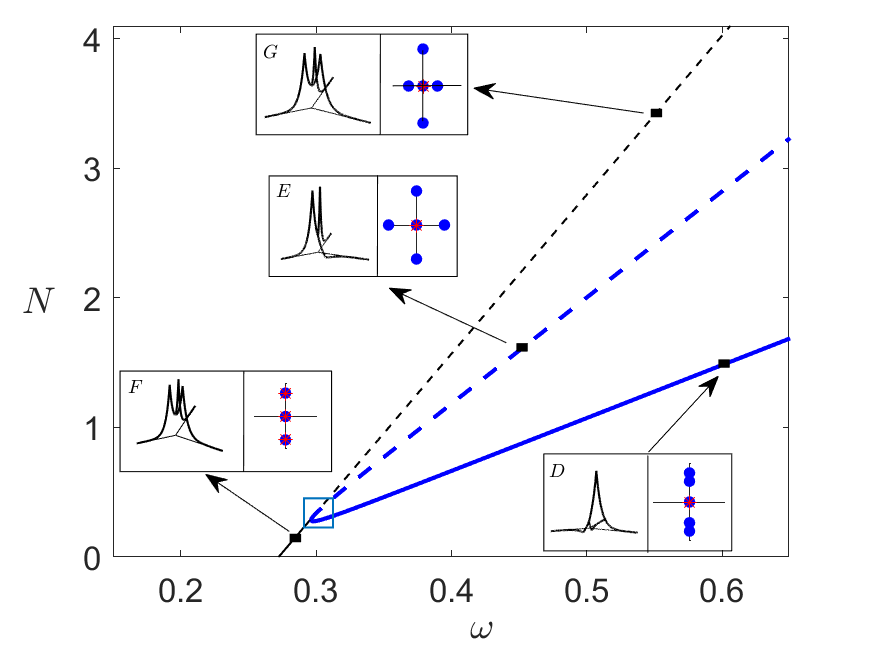}}
	\subcaptionbox{\label{subfig:inset}}{\includegraphics[scale=0.5]{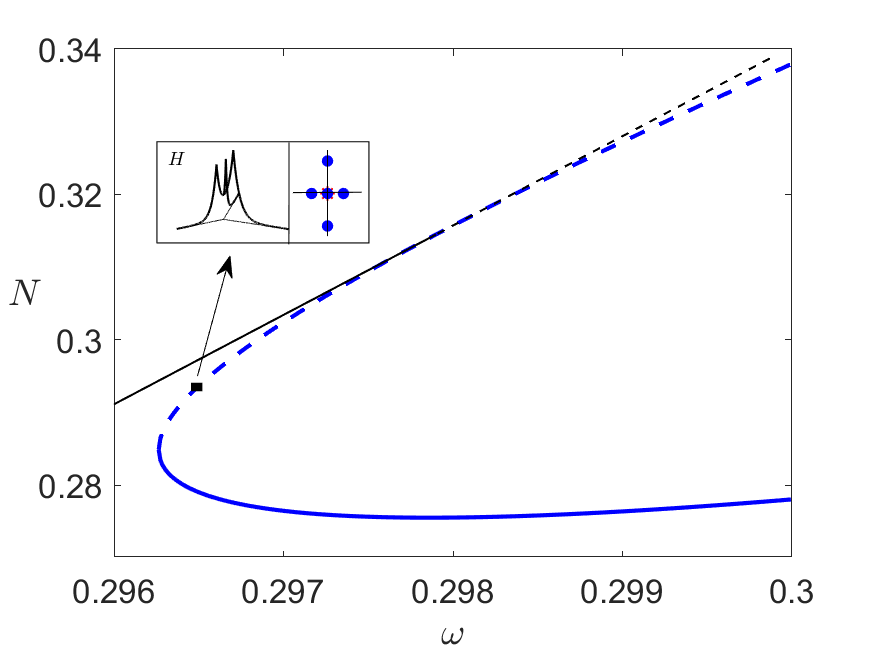}}
	\caption{Bifurcation diagrams of the equilibrium solutions discussed in Section \ref{eqm}. Plotted are the squared norms of the solutions as a function of $\omega$ for $a = 3$.  Panel (c) is a zoom-in on the small square box in panel (b), showing the solution profiles and their corresponding spectra in the complex plane.}
	\label{fig:solABC}
\end{figure}

We now analyze the equilibrium solutions of the system \eqref{sys2}. Due to the gauge invariance of the system \eqref{nls}, the equilibrium solutions of \eqref{sys2} satisfy the following set of nonlinear equations
\begin{subequations}
	\begin{align} 
	0 &= \Gamma \left(\omega_0 - \omega\right) c_0 + c_0^3 + 3 c_0 c_1^2 + \frac{3}{\sqrt{2}} c_2 c_1^2 + 3 c_0 c_2^2 - \frac{1}{\sqrt{2}} c_2^3, \label{sys5a} \\
	0 &= \Gamma \left(\omega_1 - \omega\right) c_1 + 3 c_1 c_0^2 + \frac{3}{2} c_1 c_2^2+ \frac{6}{\sqrt{2}} c_0 c_1 c_2  + \frac{3}{2} c_1^3 , \label{sys5b} \\
	0 &= \Gamma \left(\omega_1 - \omega\right) c_2 + \frac{3}{\sqrt{2}} c_0 c_1^2 + 3  c_2c_0^2 + \frac{3}{2}  c_2c_1^2 - \frac{3}{\sqrt{2}} c_0 c_2^2 + \frac{3}{2} c_2^3. \label{sys5c}
	\end{align} 
	\label{sys5}
\end{subequations}

In the following, we solve Eqs.~\eqref{sys5} for $c_0$, $c_1$, and $c_2$. All possible solutions can be categorized into three cases, which are illustrated in Fig.~\ref{fig:sketch} for the general case $a \gg 1$. An actual plot of the bifurcation diagrams for $a = 3$ is provided in Fig.~\ref{fig:solABC}. In the figure, we also describe the stability of the solutions by plotting their eigenvalues in the complex plane, which will be discussed in Section \ref{stability}.

\subsubsection{Case $c_0 = 0$} 
\label{sol1}

In this case, we consider equilibria in the subspace spanned by $u_1$ and $u_2$. When $c_0 = 0$, Eqs.~\eqref{sys5} reduce to the system
\begin{subequations}
	\begin{align} 
	\frac{3 c_1^2 c_2}{\sqrt{2}} - \frac{c_2^3}{\sqrt{2}} &= 0, \label{sys6a} \\
	\Gamma c_1 (\omega_1 - \omega) + \frac{3 c_1^3}{2} + \frac{3 c_1 c_2^2}{2} &= 0, \label{sys6b} \\
	\Gamma c_2 (\omega_1 - \omega) + \frac{3 c_1^2 c_2}{2} + \frac{3 c_2^3}{2} &= 0. \label{sys6c}
	\end{align} 
	\label{sys6}
\end{subequations}

Solving \eqref{sys6a} for $c_1$ and substituting into \eqref{sys6b} and \eqref{sys6c} yields the following solutions
\[
S_{1,2} = \left(0, \sqrt{\frac{1}{6} \Gamma (\omega - \omega_1)}, \pm \sqrt{\frac{1}{2} \Gamma (\omega - \omega_1)} \right),
\]
and
\[
S_3 = \left(0, \sqrt{\frac{2}{3} \Gamma (\omega - \omega_1)}, 0 \right).
\]
Although these equilibria are distinct, Remark \ref{rem1} shows that they correspond to the same solution when substituted into the ansatz \eqref{ansatz}. For $a = 3$, this solution is represented by the thin red curve in Fig.~\ref{subfig:solABC_u1u2}, which includes points $A$ and $B$.

\subsubsection{Case $c_1 = 0$} 
\label{sol2}

When $c_1 = 0$, Eqs.~\eqref{sys5} reduce to
\begin{subequations}
	\begin{align} 
	c_0^3 + 3 c_0 c_2^2 - \frac{c_2^3}{\sqrt{2}} + c_0 \Gamma (\omega_0 - \omega) &= 0, \label{sys7a} \\
	3 c_0^2 c_2 - \frac{3}{\sqrt{2}} c_0 c_2^2 + \frac{3}{2} c_2^3 + c_2 \Gamma (\omega_1 - \omega) &= 0. \label{sys7b}
	\end{align} 
	\label{sys7}
\end{subequations}

Solving \eqref{sys7b} for $c_2$, we obtain either $c_2 = 0$ or
\[
c_2 = \frac{1}{6} \left(3 \sqrt{2} c_0 - \sqrt{6} \sqrt{4 \Gamma \omega - 4 \Gamma \omega_1 - 9 c_0^2} \right).
\]
For $c_2 = 0$, the solution is
\[
S_4 = \left(\sqrt{\Gamma (\omega - \omega_0)}, 0, 0 \right).
\]
This solution represents a continuation of $u_0$, which, as we will see later, undergoes a symmetry-breaking bifurcation. For $a = 3$, this solution corresponds to the thin black curve in Fig.~\ref{subfig:solABC_u0}, which includes points $F$ and $G$.

For the non-zero $c_2$ case, substituting into \eqref{sys7a} yields
\begin{equation} 
9 \sqrt{3} c_0^2 \sqrt{4 \Gamma (\omega - \omega_1) - 9 c_0^2} + \sqrt{3} \Gamma (\omega_1 - \omega) \sqrt{4 \Gamma (\omega - \omega_1) - 9 c_0^2} - 9 \Gamma c_0 (\omega_0 - \omega_1) = 0,
\label{im1}
\end{equation}
which can be rewritten as the polynomial
\[
729 c_0^6 - 486 \Gamma (\omega - \omega_1) c_0^4 + 27 \Gamma^2 \left(3 \omega^2 - 2 \omega_1 (3 \omega + \omega_0) + \omega_0^2 + 4 \omega_1^2 \right) c_0^2 - 4 \Gamma^3 (\omega - \omega_1)^3 = 0.
\]
This is a cubic polynomial in $c_0^2$, and using Cardan's method \cite{nickalls1993new}, we have the following cases.

\begin{enumerate}
	\item[(1)] Within the interval $\omega_1 < \omega < \omega_{t_0}$, where $\omega_{t_0} = \sqrt{1 + 2/\sqrt{3}} (\omega_0 - \omega_1) + \omega_1$, there is only one solution
	\[
	S_5 = \left(c_0, 0, \frac{1}{6} \left(3 \sqrt{2} c_0 - \sqrt{6} \sqrt{4 \Gamma \omega - 4 \Gamma \omega_1 - 9 c_0^2} \right) \right),
	\]
	where
	\[
	c_0 = -\left(\frac{2}{9} \Gamma (\omega - \omega_1) + \frac{\sqrt[3]{-\sqrt{Y_1^2 - h_1^2} - Y_1} + \sqrt[3]{\sqrt{Y_1^2 - h_1^2} - Y_1}}{9 \sqrt[3]{2}} \right)^{1/2},
	\]
	with
	\[
	\begin{aligned}
	Y_1 &= -2 \Gamma^3 (\omega - \omega_1) \left(\omega^2 - 2 \omega \omega_1 - 3 \omega_0^2 + 6 \omega_0 \omega_1 - 2 \omega_1^2 \right), \\
	h_1 &= 2 \left(\Gamma^2 (\omega - \omega_0) (\omega + \omega_0 - 2 \omega_1) \right)^{3/2}.
	\end{aligned}
	\]

	\item[(2)] Within the interval $\omega > \omega_{t_0}$, there are three solutions
	\[
	\begin{aligned}
	c_0 &= \frac{\sqrt{2 \Gamma}}{3} \sqrt{\omega - \omega_1 + G_1 \cos \left(\theta_1 \right)}, \\
	c_0 &= -\frac{\sqrt{2 \Gamma}}{3} \sqrt{\omega - \omega_1 - G_1 \sin \left(\theta_1 + \frac{\pi}{6} \right)}, \\
	c_0 &= \frac{\sqrt{2 \Gamma}}{3} \sqrt{\omega - \omega_1 - G_1 \sin \left(\frac{\pi}{6} - \theta_1 \right)},
	\end{aligned}
	\]
	where
	\[
	\cos 3\theta_1 = \frac{\Gamma^3 (\omega - \omega_1) \left(\omega^2 - 2 \omega \omega_1 - 3 \omega_0^2 + 6 \omega_0 \omega_1 - 2 \omega_1^2 \right)}{\left(\Gamma^2 (\omega - \omega_0) (\omega + \omega_0 - 2 \omega_1) \right)^{3/2}},
	\]
	and $G_1 = \sqrt{(\omega - \omega_0) (\omega + \omega_0 - 2 \omega_1)}$. These solutions are denoted by $S_6$, $S_7$, and $S_8$, respectively.
\end{enumerate}

For $a = 3$, the solution $S_7$ is represented by the thick blue curve in Fig.~\ref{subfig:solABC_u1u2}. It meets the curve of $S_5$ at $\omega_{t_0}$, while the solutions corresponding to $S_6$ and $S_8$ bifurcate from the curve of $S_4$ in Fig.~\ref{subfig:solABC_u0}.

\subsubsection{Case $c_0, c_1, c_2 \neq 0$} 
\label{sol3}

In this case, we consider equilibria where none of the coefficients $c_0$, $c_1$, or $c_2$ vanish. Solving \eqref{sys5b} for $c_0$ yields
\[
c_0 = \frac{1}{6} \left(\sqrt{6} \sqrt{2 \Gamma (\omega - \omega_1) - 3 c_1^2} - 3 \sqrt{2} c_2 \right).
\]
Substituting this into \eqref{sys5a} and \eqref{sys5c} results in
\[
\begin{aligned}
81 c_2^3 + 18 \Gamma c_2 (\omega_0 - \omega_1) - 27 c_2^2 K + 2 \Gamma (2 \omega - 3 \omega_0 + \omega_1) K + 3 c_1^2 \left(-5 K - 9 c_2 \right) &= 0, \\
\left(c_1^2 - 3 c_2^2 \right) \left(K - 3 c_2 \right) &= 0,
\end{aligned}
\]
where $K = \sqrt{6 \Gamma (\omega - \omega_1) - 9 c_1^2}$. 

We focus on the case $c_1^2 = 3 c_2^2$, as the case $K - 3 c_2 = 0$ yields solutions already obtained in Section \ref{sol1} (i.e., the case $c_0 = 0$). This leads to
\begin{equation} 
\Gamma (2 \omega - 3 \omega_0 + \omega_1) \sqrt{6 \Gamma (\omega - \omega_1) - 27 c_2^2} - 36 c_2^2 \sqrt{6 \Gamma (\omega - \omega_1) - 27 c_2^2} + 9 \Gamma c_2 (\omega_0 - \omega_1) = 0.
\label{im2}
\end{equation}

Using a procedure similar to that in Subsection \ref{sol2}, we obtain the solutions for $c_2$ satisfying \eqref{im2} as follows

\begin{enumerate}
	\item[(1)] Within the interval $\omega_1 < \omega < \omega_{t_0}$, there is only one solution
	\[
	S_9 = \left(\frac{1}{6} \left(\sqrt{6} \sqrt{2 \Gamma (\omega - \omega_1) - 9 c_2^2} - 3 \sqrt{2} c_2 \right), \sqrt{3} c_2, c_2 \right),
	\]
	where
	\[
	c_2 = \left(\frac{1}{18} \Gamma (2 \omega - \omega_0 - \omega_1) + \frac{\sqrt[3]{-\sqrt{Y_2^2 - h_2^2} - Y_2} + \sqrt[3]{\sqrt{Y_2^2 - h_2^2} - Y_2}}{18 \sqrt[3]{2}} \right)^{1/2},
	\]
	with
	\[
	\begin{aligned}
	Y_2 &= -\Gamma^3 \left(2 \omega^3 - 3 \omega_1 \left(3 \omega^2 + 2 \omega \omega_0 + \omega_0^2 \right) + 3 \omega^2 \omega_0 + 6 \omega_1^2 (2 \omega + \omega_0) + \omega_0^3 - 6 \omega_1^3 \right), \\
	h_2 &= 2 \left(\Gamma^2 (\omega - \omega_1) (\omega + \omega_0 - 2 \omega_1) \right)^{3/2}.
	\end{aligned}
	\]

	\item[(2)] Within the interval $\omega > \omega_{t_0}$, there are three solutions
	\[
	\begin{aligned}
	c_2 &= \frac{\sqrt{\Gamma}}{3 \sqrt{2}} \sqrt{2 \omega - \omega_0 - \omega_1 + G_2 \cos \left(\theta_2 \right)}, \\
	c_2 &= -\frac{\sqrt{2 \Gamma}}{3} \sqrt{2 \omega - \omega_0 - \omega_1 - G_2 \sin \left(\theta_2 + \frac{\pi}{6} \right)}, \\
	c_2 &= \frac{\sqrt{2 \Gamma}}{3} \sqrt{2 \omega - \omega_0 - \omega_1 - G_2 \sin \left(\frac{\pi}{6} - \theta_2 \right)},
	\end{aligned}
	\]
	where
	\[
	\cos 3\theta_2 = \frac{\Gamma^3 \left(2 \omega^3 - 3 \omega_1 \left(3 \omega^2 + 2 \omega \omega_0 + \omega_0^2 \right) + 3 \omega^2 \omega_0 + 6 \omega_1^2 (2 \omega + \omega_0) + \omega_0^3 - 6 \omega_1^3 \right)}{2 \left(\Gamma^2 (\omega - \omega_1) (\omega + \omega_0 - 2 \omega_1) \right)^{3/2}},
	\]
	and $G_2 = 2 \sqrt{(\omega - \omega_1) (\omega + \omega_0 - 2 \omega_1)}$. These solutions are denoted by $S_{10}$, $S_{11}$, and $S_{12}$, respectively.

	For $a = 3$, we find using Remark \ref{rem1} that $S_{j}$, $j = 10, 11, 12$, yield the same solutions as $S_{j}$, $j = 6, 7, 8$, presented in Subsection \ref{sol2}. This is also illustrated in the sketch Fig.~\ref{fig:sketch}. These solutions are represented by the blue curves in Fig.~\ref{fig:solABC}.
\end{enumerate}

\section{Stability and Dynamics Near the Nonlinear Bound States}
\label{stability}

After obtaining all the equilibrium solutions, we now analyze their stability by solving the corresponding linear eigenvalue problem. Using the linearization ansatz
\[
c_j(t) = \tilde{c}_j(t) + (x_j + i y_j) e^{\lambda t}, \quad j = 0, 1, 2,
\]
where $|x_j|, |y_j| \ll 1$ and $\tilde{c}_j(t)$ represents the equilibrium solution obtained in Section \ref{eqm}, we substitute this into \eqref{sys2} to derive the eigenvalue problem
\[
\lambda x = M x,
\]
where $x = \left(x_0, y_0, x_1, y_1, x_2, y_2\right)^T$, and $M = (m_{jk})$, $j, k = 1, 2, \dots, 6$, is the coefficient matrix. The components of $M$ are given by
\[
\begin{aligned}
m_{12} &= \Gamma (\omega - \omega_0) - \tilde{c}_0^2 - \tilde{c}_1^2 - \tilde{c}_2^2, & & m_{14} = -2 \tilde{c}_0 \tilde{c}_1 - \sqrt{2} \tilde{c}_1 \tilde{c}_2, \\
m_{16} &= -2 \tilde{c}_0 \tilde{c}_2 - \frac{\tilde{c}_1^2}{\sqrt{2}} + \frac{\tilde{c}_2^2}{\sqrt{2}}, & & m_{21} = -\Gamma (\omega - \omega_0) + 3 \tilde{c}_0^2 + 3 \tilde{c}_1^2 + 3 \tilde{c}_2^2, \\
m_{23} &= -3 m_{14}, & & m_{25} = -3 m_{16}, \\
m_{32} &= m_{14}, & & m_{34} = \Gamma (\omega - \omega_1) - \tilde{c}_0^2 - \sqrt{2} \tilde{c}_0 \tilde{c}_2 - \frac{3 \tilde{c}_1^2}{2} - \frac{\tilde{c}_2^2}{2}, \\
m_{36} &= \frac{1}{\sqrt{2}} m_{14}, & & m_{41} = -3 m_{14}, \\
m_{43} &= \Gamma (\omega - \omega_1) - 3 \tilde{c}_0^2 - 3 \sqrt{2} \tilde{c}_0 \tilde{c}_2 - \frac{9 \tilde{c}_1^2}{2} - \frac{3 \tilde{c}_2^2}{2}, & & m_{45} = -\frac{3}{\sqrt{2}} m_{14}, \\
m_{52} &= m_{16}, & & m_{54} = \frac{1}{\sqrt{2}} m_{14}, \\
m_{56} &= \Gamma (\omega - \omega_1) - \tilde{c}_0^2 + \sqrt{2} \tilde{c}_0 \tilde{c}_2 - \frac{\tilde{c}_1^2}{2} - \frac{3 \tilde{c}_2^2}{2}, & & m_{61} = -3 m_{16}, \\
m_{63} &= -\frac{3}{\sqrt{2}} m_{14}, & & m_{65} = \Gamma (\omega - \omega_1) - 3 \tilde{c}_0^2 + 3 \sqrt{2} \tilde{c}_0 \tilde{c}_2 - \frac{3 \tilde{c}_1^2}{2} - \frac{9 \tilde{c}_2^2}{2},
\end{aligned}
\]
and all other entries are zero. A solution is considered unstable if $\text{Re}(\lambda) > 0$ for some $\lambda$, and linearly stable otherwise. Since our system is Hamiltonian, linear stability requires all eigenvalues to lie on the imaginary axis, i.e., $\text{Re}(\lambda) = 0$.

We are also interested in the typical dynamics of a solution when it is unstable. We numerically solved the coupled mode equations \eqref{sys2} using a fourth-order Runge-Kutta method to investigate this. The initial condition is an unstable equilibrium perturbed by small disturbances. To present the simulation results, we substituted the time evolution of $c_j$ into Eq.~\eqref{ansatz} and plotted the resulting function. This approach provides a more informative representation than plotting $c_j(t)$ directly, as the dynamics of the field in each branch is a combination of different modes with non-zero profiles along the branches (see Fig.~\ref{fig:soln}).

In the following, we discuss the stability of each solution obtained in Section \ref{eqm}.

\subsection{Case $c_0 = 0$} 
\label{stab1}

For any of the equilibria in this case, the characteristic polynomial of the coefficient matrix $M$ is
\[
\frac{1}{3} \lambda^2 \left(\gamma + \lambda^2 + 3 \lambda^4 \right) = 0,
\]
which can be solved for the eigenvalue $\lambda$. 

Defining
\[
\beta = \Gamma^2 \left(3 \omega^2 - 8 \omega_1 (\omega + \omega_0) + 2 \omega \omega_0 + 3 \omega_0^2 + 8 \omega_1^2 \right), \quad \gamma = 4 \Gamma^4 (\omega - \omega_1)^3 (\omega_0 - \omega_1),
\]
the eigenvalues are given by
\[
\lambda^2 = \frac{-\beta \pm \sqrt{\beta^2 - 12 \gamma}}{12}.
\]
There is a change of stability, as indicated by the thin red dashed line in Fig.~\ref{subfig:solABC_u1u2}, within the interval $\omega_{t_1} < \omega < \omega_{t_2}$, where
\[
\begin{aligned}
\omega_{t_1} &= \frac{1}{9} \left(2 \left(\sqrt[3]{3 \left(\sqrt{57} + 9 \right)} + \sqrt[3]{27 - 3 \sqrt{57}} \right) \sqrt[3]{(\omega_0 - \omega_1)^3} + 3 (\omega_0 + 2 \omega_1) \right), \\
\omega_{t_2} &= 3 \omega_0 - 2 \omega_1.
\end{aligned}
\]
Within this interval, the solution is unstable.

\subsection{Case $c_1 = 0$} 
\label{stab2}

For the equilibrium $(\sqrt{\Gamma (\omega - \omega_0)}, 0, 0)$, the eigenvalues of $M$ are
\[
0, \quad \pm \Gamma \sqrt{\omega_0 - \omega_1} \sqrt{2 \omega - 3 \omega_0 + \omega_1}.
\]
Since $\omega_0 > \omega_1$, the equilibrium transitions from stable to unstable when $\omega > \omega_{t_3}$, where
\[
\omega_{t_3} = \frac{3 \omega_0 - \omega_1}{2}.
\label{wc3}
\]
The unstable solution is shown by the thin black dashed line in Fig.~\ref{subfig:solABC_u0}.

For the case where $c_0, c_2 \neq 0$, we express $\Gamma(\omega - \omega_0)$ and $\Gamma(\omega - \omega_1)$ in terms of $\tilde{c}_j$ using \eqref{sys7}. Substituting these into the coefficient matrix $M$, we obtain the eigenvalues
\[
\begin{aligned}
\lambda &= 0, \quad \pm \frac{3 \sqrt{2 \sqrt{2} \tilde{c}_0^5 \tilde{c}_2 - 5 \tilde{c}_0^4 \tilde{c}_2^2 + \sqrt{2} \tilde{c}_0^3 \tilde{c}_2^3}}{\sqrt{2} \tilde{c}_0}, \\
&\quad \pm \frac{\sqrt{-6 \sqrt{2} \tilde{c}_0^5 \tilde{c}_2 - 25 \tilde{c}_0^4 \tilde{c}_2^2 + 19 \sqrt{2} \tilde{c}_0^3 \tilde{c}_2^3 - 6 \tilde{c}_0^2 \tilde{c}_2^4 + 2 \sqrt{2} \tilde{c}_0 \tilde{c}_2^5 - \tilde{c}_2^6}}{\sqrt{2} \tilde{c}_0}.
\end{aligned}
\]
Within the interval $\omega_1 < \omega < \omega_{t_0}$, the solution is unstable. For $\omega_{t_0}<\omega < \omega_{t_3}$, there are three unstable solutions where at $\omega_{t_3}$, one of them changes its stability as shown by the thick solid and dashed blue lines in Fig.~\ref{fig:solABC}.

\subsection{Case $c_0, c_1, c_2 \neq 0$} 
\label{stab3}

Using a similar procedure, we substitute $\Gamma(\omega - \omega_0)$ and $\Gamma(\omega - \omega_1)$ obtained from \eqref{sys7} into the coefficient matrix $M$, yielding the eigenvalues
\[
\begin{aligned}
\lambda &= 0, \quad \pm \frac{3 \sqrt{2} \sqrt{\left(-\sqrt{2}\right) \tilde{c}_0^5 \tilde{c}_2 - 5 \tilde{c}_0^4 \tilde{c}_2^2 - 2 \sqrt{2} \tilde{c}_0^3 \tilde{c}_2^3}}{\tilde{c}_0}, \\
&\quad \pm \frac{\sqrt{2} \sqrt{3 \sqrt{2} \tilde{c}_0^5 \tilde{c}_2 - 25 \tilde{c}_0^4 \tilde{c}_2^2 - 38 \sqrt{2} \tilde{c}_0^3 \tilde{c}_2^3 - 24 \tilde{c}_0^2 \tilde{c}_2^4 - 16 \sqrt{2} \tilde{c}_0 \tilde{c}_2^5 - 16 \tilde{c}_2^6}}{\tilde{c}_0}.
\end{aligned}
\]
These eigenvalues are expressed in terms of $\tilde{c}_0$ and $\tilde{c}_2$. We find that they are identical to those obtained in Subsection \ref{stab2}, consistent with the fact that different equilibria correspond to the same solution $u(x)$. As in Subsection \ref{stab2}, there are two unstable solutions and one stable solution, shown by the thick blue lines in Fig.~\ref{fig:solABC}.

\section{Numerical solutions}\label{num_ori}

We directly solve the regularized version of the original equations \eqref{eqbs}, i.e., 
\begin{equation}
u^{(k)}_{xx} - \omega u^{(k)} + {u^{(k)}}^3 = 0,\quad x\neq a,
\label{reg}
\end{equation}
together with the vertex conditions \eqref{bc} and the matching conditions \eqref{match}. See also \cite{wang2009two} on different approaches to regularize the delta function $\delta(x)$. We look for localized solutions that bifurcate from the linear modes. 

Once a static solution -- let us say $\varphi^{(k)}(x)$ -- is obtained, we determine its linear stability by computing the corresponding eigenvalue problem (that can be derived similarly as that in Section \ref{stability}), i.e.,  we introduce a small variation around the static solution, expressed as $u^{(k)}(x,t) = \varphi^{(k)}(x) + (p^{(k)}(x,t) + iq^{(k)}(x,t))$, where $|p^{(k)}|,|q^{(k)}| \ll 1$ represent perturbation. After linearizing around $p^{(k)}$ and $q^{(k)}$ and separating the real and imaginary parts, we substitute $p^{(k)}(x,t) = r^{(k)}(x) e^{\lambda t}$ and $q^{(k)}(x,t) = s^{(k)}(x) e^{\lambda t}$ to arrive at the following eigenvalue problem
\begin{equation} \label{eqn:eigval_eqn}
    \lambda \begin{bmatrix}
        r^{(k)} \\ s^{(k)}
    \end{bmatrix} = \begin{bmatrix}
        0 & L_{-} \\ -L_{+} & 0
    \end{bmatrix}\begin{bmatrix}
        r^{(k)} \\ s^{(k)}
    \end{bmatrix} = \mathcal{L} \begin{bmatrix}
        r^{(k)} \\ s^{(k)}
    \end{bmatrix},\quad x\neq a,
\end{equation}
where $r^{(k)}$ and $s^{(k)}$ belong to the space of integrable functions, $L^{2}(G)$. The operators $L_{-}$ and $L_{+}$ are defined as
\begin{align}
    L_{\pm} &= \partial_{xx} - \omega + (2\pm1)\varphi^2.
\end{align}
Due to the conditions \eqref{bc} and \eqref{match}, we have the vertex conditions
\begin{eqnarray}
    \sum_{k=1}^3 \Box^{(k)}_x(0) = 0, \quad \Box^{(1)}(0) = \Box^{(2)}(0) = \Box^{(3)}(0),\label{cond1}
\end{eqnarray}
and the matching conditions 
\begin{eqnarray}
\Box^{(k)}(a^+) = \Box^{(k)}(a^-), \quad \Box^{(k)}_x(a^+) - \Box^{(k)}_x(a^-) = -\Box^{(k)}(a),\label{cond2}
\end{eqnarray}
where $\Box=r,s$. The stability of the stationary solution is dictated by the eigenvalues of the operator $\mathcal{L}$. With the same argument as that in Section \ref{stability}, due to the Hamiltonian nature of the system, if any eigenvalue $\lambda$ possesses a non-zero real part, i.e., $\text{Re}\ \lambda \neq 0$, the stationary solution is deemed unstable; otherwise, it is considered marginally stable.

To solve the static equations and the corresponding linear eigenvalue problem, we discretize them using a central finite difference. The computational domain is $0<x\leq l=30$ with the discretization $dx=0.02$. At the end $x=l$, we use Neumann boundary conditions. Using the fourth-order Runge-Kutta method, we also simulate the dynamics of unstable solutions by solving the regularized version of the governing equations \eqref{nls} with \eqref{bc}.

\section{Discussion}
\label{num}

\begin{figure}[tbhp!]
	\centering
	\setcounter{subfigure}{0}
	\subcaptionbox{}{\includegraphics[scale=0.55]{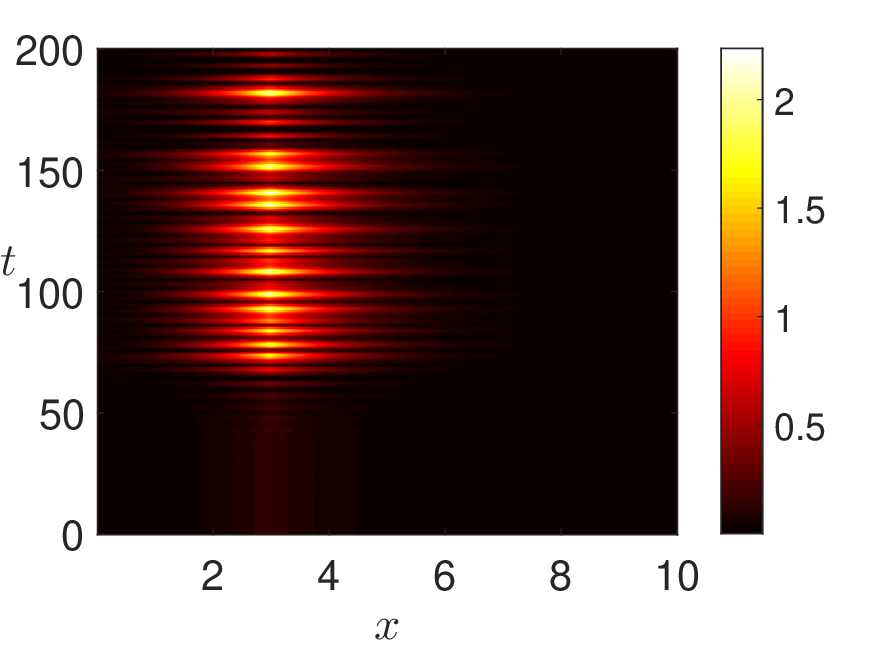}\label{subfig:A_a1}} 
	\subcaptionbox{}{\includegraphics[scale=0.55]{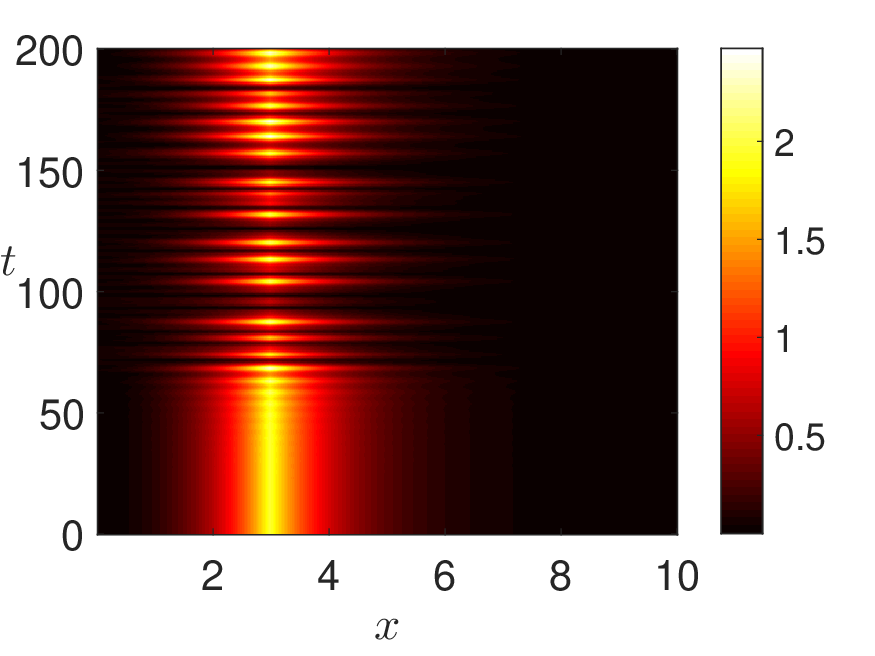}\label{subfig:A_a2}}
	\subcaptionbox{}{\includegraphics[scale=0.55]{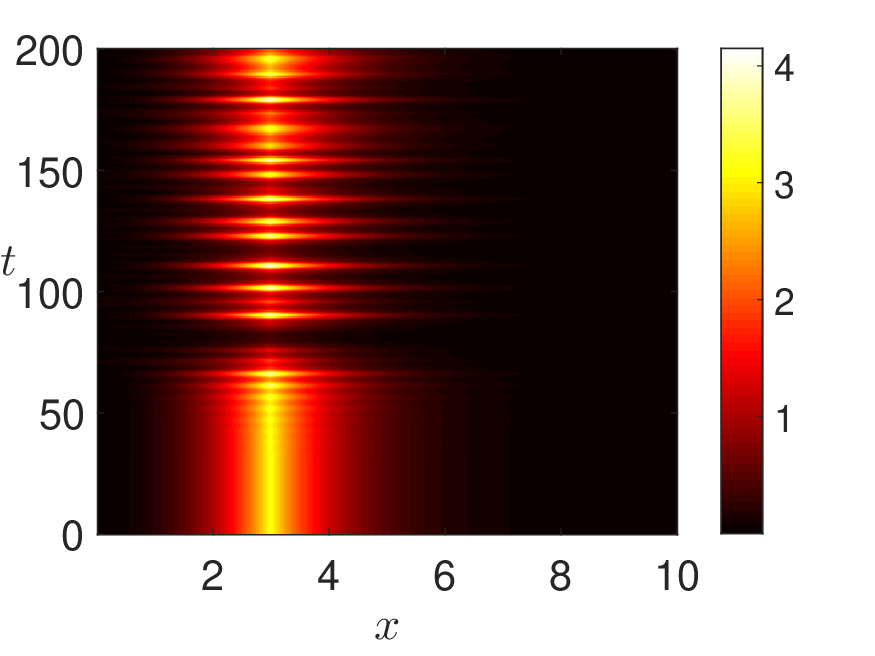}\label{subfig:A_a3}}
	\caption{Time dynamics of the unstable solution at point $A$ in Fig.~\ref{subfig:solABC_u1u2}. Shown are $|u^{(k)}|^2$, $k=1,2,3$. The system becomes chaotic.}
	\label{fig:dynamics1}
\end{figure}

\begin{figure}[htbp!]
	\centering
	\subcaptionbox{}{\includegraphics[scale=0.55]{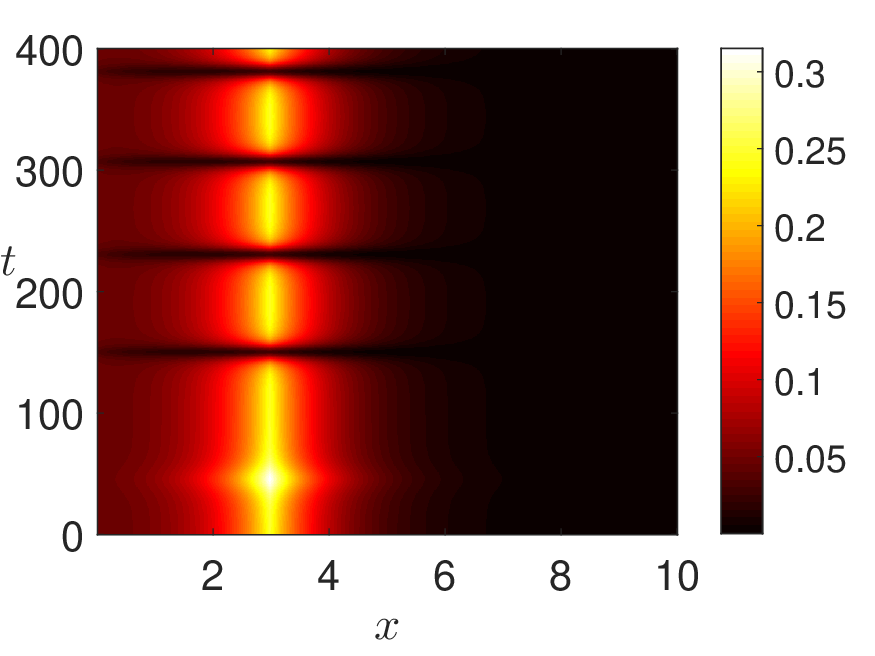}\label{subfig:H_a1_v2}}
	\subcaptionbox{}{\includegraphics[scale=0.55]{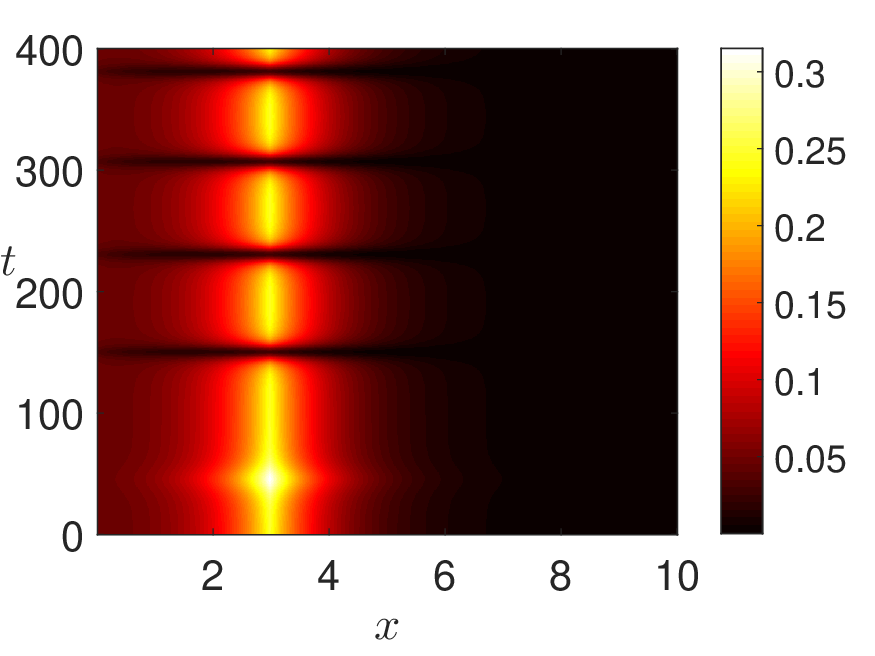}\label{subfig:H_a2_v2}}
	\subcaptionbox{}{\includegraphics[scale=0.55]{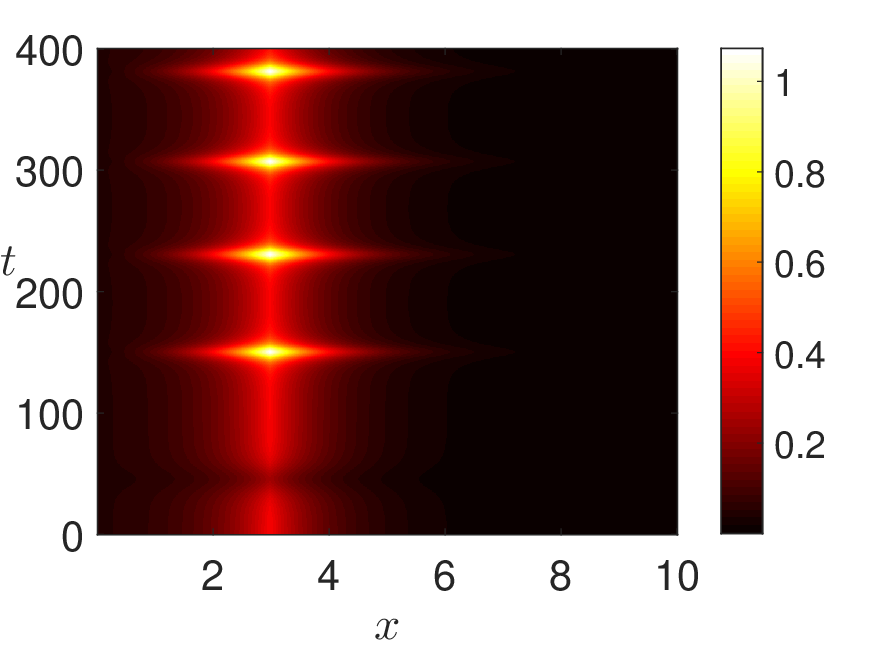}\label{subfig:H_a3_v2}}
	\caption{The same as Fig.~\ref{fig:dynamics1}, but for the unstable solution denoted by point $H$ in Fig.~\ref{subfig:inset}. The static solution is attracted to a periodic state.}
	\label{fig:dynamics2}
\end{figure}

The sketch in Fig.~\ref{fig:sketch} provides a comprehensive picture of the bifurcations of standing waves from the linear states (see Fig.~\ref{fig:soln}) for $a \gg 1$. The actual plots from solving the coupled-mode equations are shown in Figs.~\ref{fig:solABC}. The thin red curve in Fig.~\ref{subfig:solABC_u1u2} shows the existence curve of a nonlinear state bifurcating from the eigenfrequency $\omega_1$. The solution is stable for values of $\omega$ close to the bifurcation point. As $\omega$ varies, there is an interval where the solution becomes unstable through a Hamiltonian-Hopf bifurcation. In this interval, two pairs of eigenvalues with non-zero real parts indicate oscillatory instability.

We considered the unstable solution denoted by point $A$ in Fig.~\ref{subfig:solABC_u1u2}. Its dynamics are depicted in Fig.~\ref{fig:dynamics1}. Around $t \approx 50$, the oscillatory nature of the instability becomes evident, eventually leading to chaotic dynamics. 

The dashed blue curve in Fig.~\ref{subfig:solABC_u1u2} shows a bifurcation of another family of solutions from $\omega_1$. The solution denoted by point $C$ can be seen as the continuation of the linear state $u_2$ (see Fig.~\ref{fig:soln}). Our analysis reveals that the nonlinear continuation is always unstable in its existence region. The instability is due to a pair of real eigenvalues, indicating exponential instability. We simulated this unstable solution and observed that the long-time dynamics are also chaotic. The only visual difference with the oscillatory instability in Fig.~\ref{fig:dynamics1} is in the initial dynamics, where the instability manifests as a continuous increase or decrease of the fields rather than oscillations.

\begin{figure}[tbhp!]
	\centering
	\setcounter{subfigure}{0}
	\subcaptionbox{}{\includegraphics[scale=0.55]{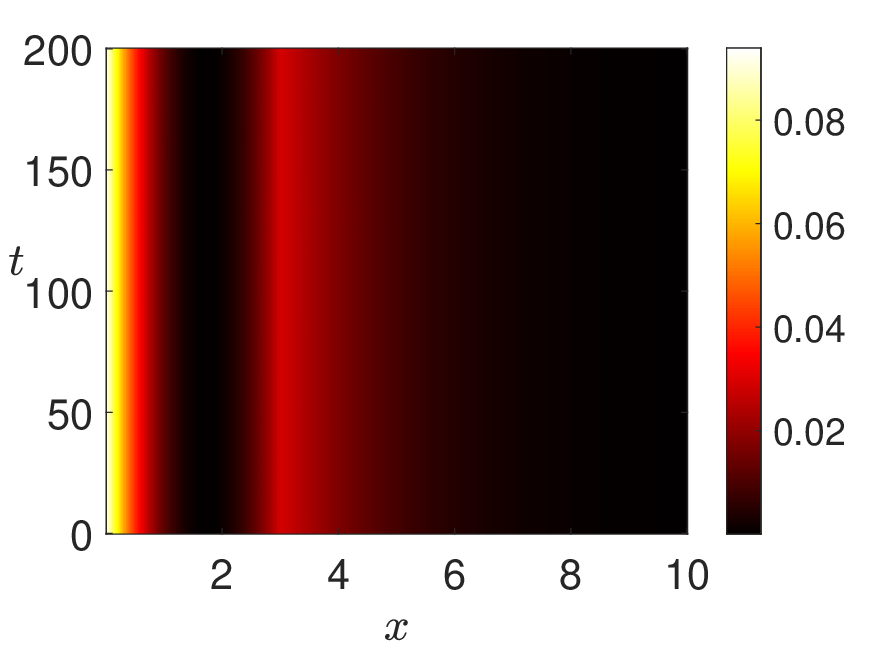}\label{subfig:D_a1}} 
	\subcaptionbox{}{\includegraphics[scale=0.55]{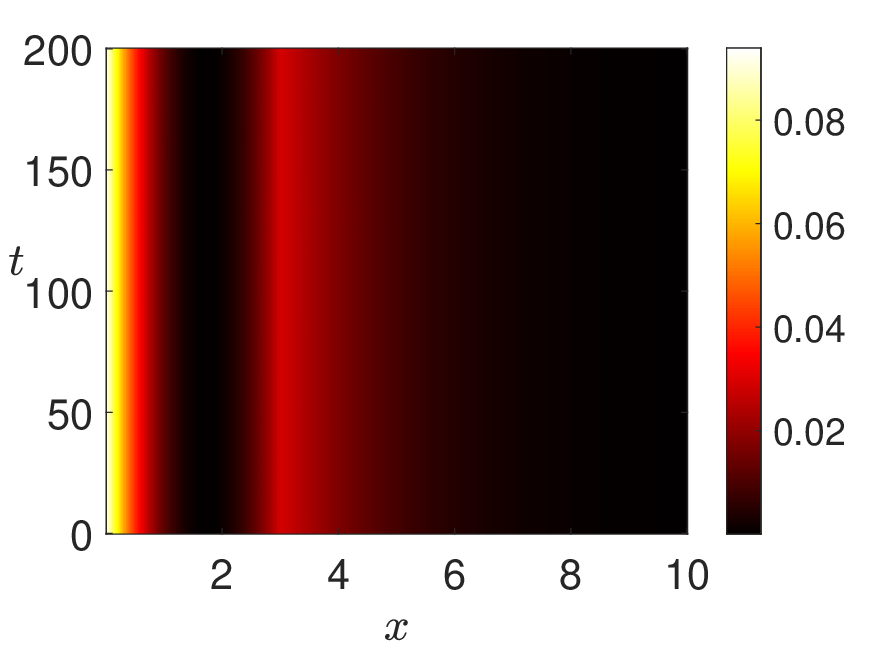}\label{subfig:D_a2}}
	\subcaptionbox{}{\includegraphics[scale=0.55]{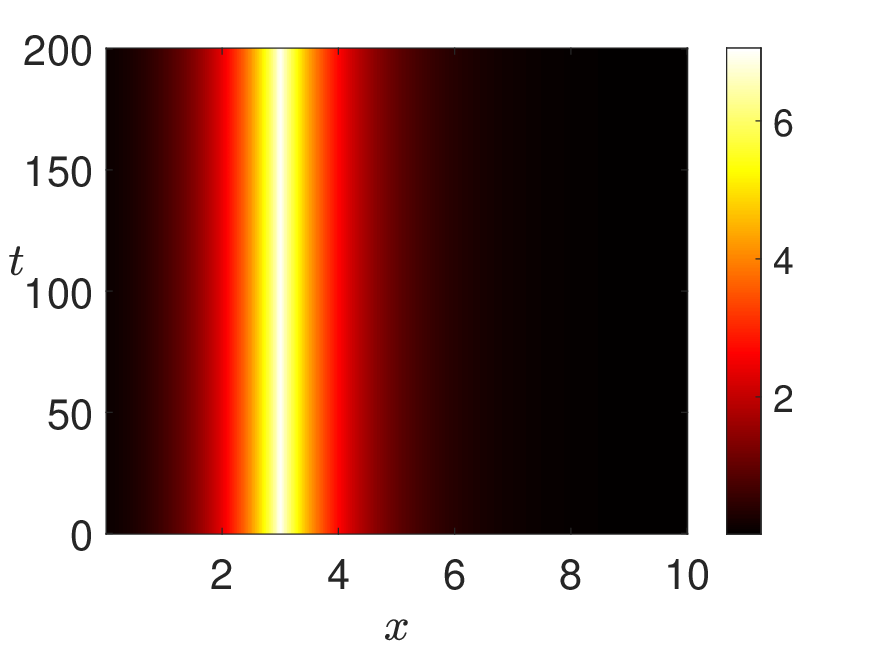}\label{subfig:D_a3}}
	\caption{Time dynamics of the stable solution at point $D$ in Fig.~\ref{subfig:solABC_u1u2}. Shown are $|u^{(k)}|^2$, $k=1,2,3.$}
	\label{fig:dynamics3}
\end{figure}

Figure \ref{subfig:solABC_u0} presents one of the main results of the paper: the bifurcation of the ground state $u_0$. Near the bifurcation point $\omega_0$, the nonlinear state is stable. As $\omega$ increases further, there is a threshold value $\omega_{t_3}$ \eqref{wc3} where the symmetric state becomes (exponentially) unstable. At this point, two asymmetric states bifurcate in a supercritical-like manner for $\omega > \omega_{t_3}$, but the bifurcating solution is also unstable. Additionally, two asymmetric states bifurcate in a subcritical manner for $\omega < \omega_{t_3}$. Figure \ref{subfig:inset} zooms in on the area around the subcritical bifurcation. Near $\omega_{t_3}$, the asymmetric solutions arise due to the interaction of the modes $u_0$ with $u_1$ and $u_2$. We also simulated the unstable solution denoted by point $H$ in Fig.~\ref{subfig:inset}. The dynamics are plotted in Fig.\ \ref{fig:dynamics2}. Unlike the previous unstable case, here the solution approaches a periodic solution as $t$ increases. We also present the time dynamics of a stable asymmetric solution, denoted by point $D$ in Fig.\ \ref{fig:dynamics3}.

\begin{figure}[htbp!]
	\centering
	\subcaptionbox{\label{figa}}{\includegraphics[scale=0.55]{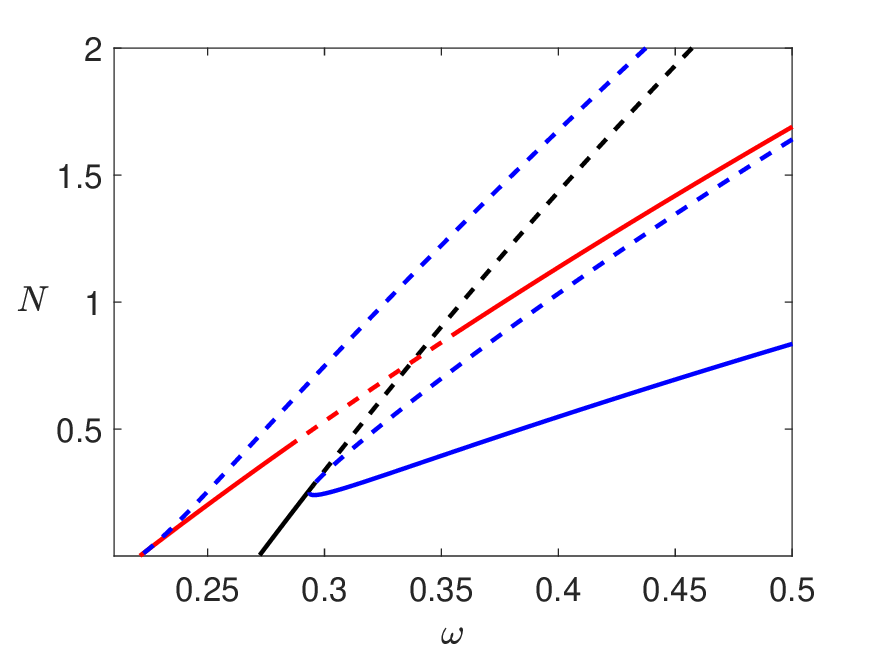}}
    \subcaptionbox{\label{figb}}{\includegraphics[scale=0.55]{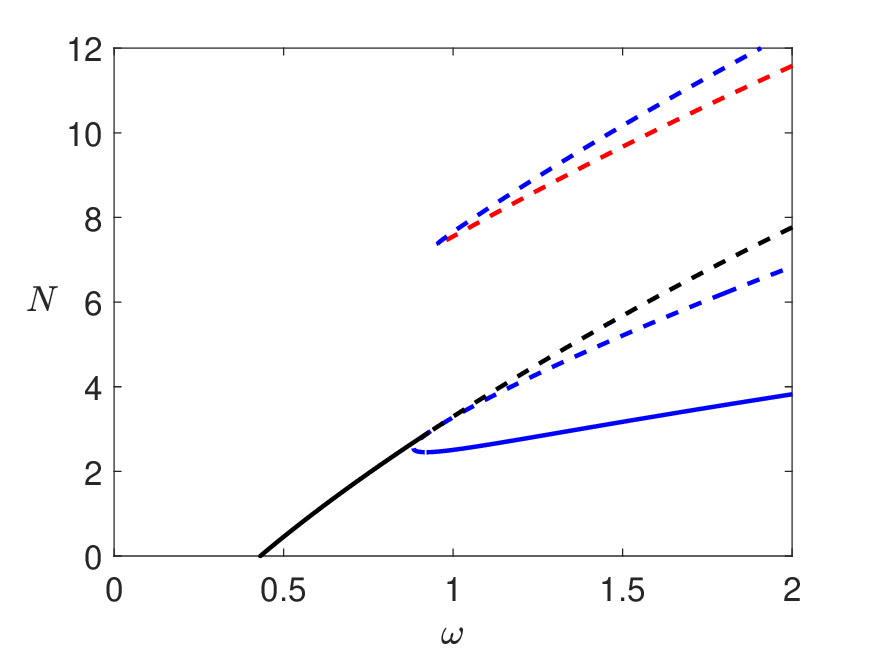}}
	\caption{Bifurcation diagrams of the equilibrium solutions obtained from the original equations \eqref{eqbs}. Plotted are the squared norms of the solutions as a function of $\omega$ for $a = 3$ (left) and $a=0.9$ (right). Note the close resemblance of the left diagrams with those from the coupled mode approximations in Fig.\ \ref{fig:solABC}.}
	\label{fig:ori}
\end{figure}

The bifurcation diagrams of the localized solutions from the original system, as discussed in Section \eqref{num_ori}, are presented in Fig.\ \ref{fig:ori}. Notably, our numerical results for $a\gg1$, i.e., Fig.\ \ref{figa}, perfectly agree with the predictions from the coupled-mode approximations shown in Fig.\ \ref{fig:solABC}. This strong correspondence shows the accuracy and validity of the coupled-mode approach in capturing the essential features of the system. The dynamics of unstable solutions are also similar to those plotted in Figs.\ \ref{fig:dynamics1} and \ref{fig:dynamics2} and hence are not shown here.

As a final remark, we note that the symmetry-breaking bifurcation reported above is clearly due to the interaction of several modes, as seen in the ansatz \eqref{ansatz}. Such an ansatz is only possible for $a > 1$ (see Fig.~\ref{fig:a_vs_omega}). When $a < 1$, only the linear symmetric state $u_0$ exists. However, our numerics in Fig.\ \ref{figb} show that the positive state $u_0$ still undergoes a symmetry-breaking bifurcation at large enough $\omega$. As $\omega_1$ disappears, the continuation of the linear modes $u_1$ and $u_2$ is now connected through a turning point. 

\section{Conclusion and future works}
\label{concl}

In this paper, we have investigated the bifurcations of nonlinear states from their linear counterparts and analyzed their stability. By deriving a finite-dimensional dynamical system approximation using the coupled mode reduction method, we presented novel results on symmetry-breaking bifurcations of the positive definite solutions, which are the ground states. The bifurcating asymmetric states were shown to be unstable, although one of them regained stability after a turning point. We also provided a detailed discussion of the continuations of excited states. Additionally, we presented typical time dynamics of the unstable solutions, where we observed either chaotic dynamics or periodic states. 

It is interesting to explore the origin of these dynamics, using the studies of, e.g., \cite{goodman2011hamiltonian,goodman2017bifurcations} for triple-well potentials on the real line. Another important problem is extending the present study to star graphs with many edges, aiming to understand the general picture of symmetry-breaking bifurcations in such systems. This will also be reported in future publications.

The instability of the symmetric state cannot be explained by coupled-mode approximations when $a<1$. Its origin will be studied in the future. Furthermore, in the limit $a \to 0$, the work of \cite{adami2012stationary} shows that a symmetric trapped soliton at the vertex with a $\delta$-interaction is the ground state for any solution norm. The relation between our work and the results of \cite{adami2012stationary} is an open problem.

\section*{Acknowledgments}
R.R.\ is funded by the Directorate of Research and Development, Universitas Indonesia, under Hibah PUTI 2022 (Grant No.\ NKB-1461/UN2.RST/HKP.05.00/2022). HS acknowledged support by Khalifa University through a Competitive Internal Research Awards Grant (No.\ 8474000413/CIRA-2021-065) and Research \& Innovation Grants (No.\ 8474000617/RIG-S-2023-031 and No.\ 8474000789/RIG-S-2024-070). The authors contribute equally to the manuscript. We also thank Rudy Kusdiantara for assistance in converting the figure file from PDF to EPS, which resolved persistent submission issues and enabled a successful arXiv submission after several unsuccessful attempts.

\bibliographystyle{elsarticle-num}
	\bibliography{references}

\end{document}